\documentstyle[aps,prb,twocolumn]{revtex}

\begin{document}

\input epsf.sty
\twocolumn[\hsize\textwidth\columnwidth\hsize\csname %
@twocolumnfalse\endcsname

\draft
\widetext

\title
{
Static magnetic correlations near the insulating-superconducting phase boundary in La$_{2-x}$Sr$_{x}$CuO$_{4}$
}

\author
{
M. Fujita and K. Yamada
}

\address
{
Institute for Chemical Research, Kyoto University, Gokasho, 
Uji 610-0011, Japan
}

\author
{
H. Hiraka\footnote{Now at the Institute for Material Research, Tohoku University, Katahira, Sendai 980-0821}
}

\address
{
Department of Physics, Tohoku University, Sendai 980-8578, Japan
}

\author
{
P. M. Gehring and S. H. Lee\footnote{Also at the University of Maryland, College Park, MD 20742
}}

\address
{
National Institute of Standards and Technology, NCNR, Gaithersburg, Maryland 20889
}

\author
{
S. Wakimoto\footnote{Also at the Department of Physics, Massachusetts Institute of Technology, Cambridge, MA 02139} 
and G. Shirane
}

\address
{
Department of Physics, Brookhaven National Laboratory, Upton, NY 11973
}

\date{\today}

\maketitle




\begin{abstract}
An elastic neutron scattering study has been performed on several single crystals of La$_{2-x}$Sr$_{x}$CuO$_{4}$ 
for {\it x} near the lower critical concentration {\it x$_{c}$} for superconductivity.  
In the insulating spin-glass phase ({\it x} = 0.04 and 0.053), the previously reported one-dimensional spin modulation along the orthorhombic {\it b}-axis is confirmed. 
Just inside the superconducting phase ({\it x} = 0.06), however, two pairs of incommensurate magnetic peaks are additionally observed corresponding to the spin modulation parallel to the tetragonal axes. 
These two types of spin modulations with similar incommensurabilities coexist near the boundary.
The peak-width $\kappa$ along spin-modulation direction exhibits an anomalous maximum in the superconducting phase near {\it x$_{c}$}, where the incommensurability $\delta$ monotonically increases upon doping across the phase boundary. 
These results are discussed in connection with the doping-induced superconducting phase transition. 
\end{abstract}


\pacs{PACS numbers: 74.72.Dn, 75.30.Fv, 75.50.Ee}

\phantom{.}
]
\narrowtext


\section{Introduction}
\label{sec_intro}

The intimate connection between the novel superconductivity and magnetism found in the high-{\it T$_{c}$} cuprates is believed to be fundamental to the underlying superconducting mechanism.~\cite{Kastner_98} 
Extensive neutron scattering measurements of La$_{2-x}$Sr$_{x}$CuO$_{4}$ (LSCO) have revealed the doping dependence of the low-energy magnetic spin fluctuations over a wide range of doping.~\cite{Yamada_98} 
More recent studies have focused attention on the static or quasi-static magnetic ordering that coexists or phase-separates with the superconductivity.~\cite{Tranquada_95}  Incommensurate (IC) elastic peaks were first observed in La$_{1.48}$Nd$_{0.4}$Sr$_{0.12}$CuO$_{4}$ around the ($\pi$,$\pi$) position, which corresponds to (1/2,1/2,0) in tetragonal notation as shown in the right inset of Fig. 1(b). 
Similar peaks have also been observed in  La$_{1.88}$Sr$_{0.12}$CuO$_4$~\cite{Suzuki_98,Kimura_99} and La$_2$CuO$_{4.12}$,~\cite{YoungLee_99} although the peak positions are slightly shifted towards a more rectangular arrangement compared to the square geometry found in La$_{1.48}$Nd$_{0.4}$Sr$_{0.12}$CuO$_{4}$.~\cite{Tranquada_96} 
In these superconducting systems, since the spin modulation vector is parallel~\cite{Tranquada_96} or almost parallel to the Cu-O bond~\cite{YoungLee_99,Kimura_00}, we use the term parallel spin modulation or correlations.

Recently, another type of IC magnetic order was discovered by Wakimoto {\it et al}. in the insulating spin-glass phase at {\it x} = 0.03, 0.04 and 0.05.~\cite{Waki_99_1,Waki_00_1} 
The IC peaks observed at these Sr concentrations are located at the diagonal positions depicted in the reciprocal lattice diagram shown in the left inset of Fig. 1(b). 
Another important discovery is that the spin modulation is observable only along the orthorhombic {\it b}-axis. 
Such a one-dimensional nature for the spin correlations is consistent with a stripe-like ordering of the holes in the CuO$_{2}$ planes. 
More recently, similar diagonal IC peaks were found in LSCO samples with {\it x} = 0.024, just above the critical concentration for three-dimensional N$\acute{e}$el order.~\cite{Matsuda_00} 
Therefore, the diagonal spin density modulation is considered to be an intrinsic property of the entire insulating spin-glass region, and stands in stark contrast to the parallel spin modulation observed in the superconducting phase. 
These results strongly suggest that a drastic change takes place in the spin modulation vector, from diagonal to parallel, near the lower critical concentration for superconductivity $x_{c}$$\approx$0.055.

Important questions to be resolved are how the change in the spin density modulation occurs and how it is related to the insulating-to-superconducting phase transition. 
In addition, the nature of the doping-induced superconducting phase transition itself is a key issue related to the position of quantum critical point in the high-{\it T$_{c}$} cuprate phase diagram. 
To shed more light on these questions, we have carried out a systematic series of elastic neutron scattering experiments on single crystals of LSCO with {\it x} = 0.053, 0.056, 0.06 and 0.07 that spans the insulating-superconducting phase boundary. 
Quantitative analyses are also presented here on data obtained from other samples 
\linebreak
\vspace{-0.5cm}
\begin{figure}[H]
\centerline{\epsfxsize=2.65in\epsfbox{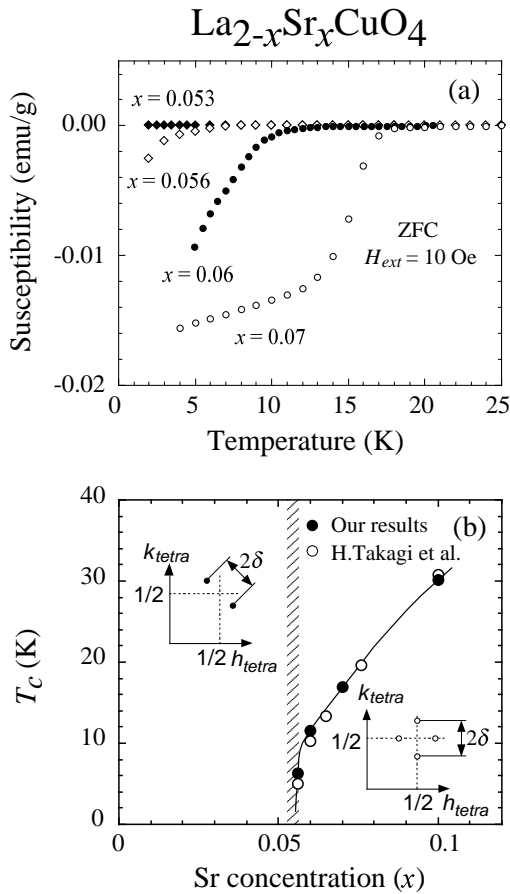}}
\vspace{0.5cm}
\caption
{(a) Magnetic susceptibility, measured at 10 Oe, for zero-field-cooled single crystals of La$_{2-x}$Sr$_{x}$CuO$_{4}$ with {\it x} = 0.053, 0.056, 0.06, and 0.07. (b)  Doping dependence of the superconducting transition temperature {\it T$_{c}$} determined by susceptibility measurements. The shaded line depicts the boundary between insulating and superconducting phases at lowest temperature. Insets show the magnetic IC peak positions in the insulating (left) or superconducting (right) phase.}
\end{figure}
\vspace{0in}
\noindent
with {\it x} = 0.03 and 0.04 which were not shown in detail in Ref. 10. 
From all of these data we can confirm that the appearance of the parallel spin correlations coincides with that of the superconductivity. 
On the other hand, the diagonal spin correlations persist into the superconducting state near the phase boundary, where they also exhibit an anomalous broadening of the peak-width. 
The incommensurability $\delta$, for both the diagonal and parallel peaks, varies monotonously across the phase boundary.

The preparation and characterization of our LSCO single crystals and experimental details are described in Section II of this paper. 
Data from the neutron scattering measurements taken on single crystals in both the insulating spin-glass and superconducting phases are introduced in Section III. 
Finally, in Section IV we discuss the nature of the doping-induced superconducting phase transition.

\section{IEXPERIMENTAL DETAILS}

\subsection{Sample Preparation and Characterization}

A series of single crystals with {\it x} = 0.053, 0.056, and 0.07 were grown using a traveling-solvent floating-zone method.~\cite{Tanaka_89} 
By utilizing large focusing mirrors in our new furnaces we are able to keep the temperature gradient around the molten zone stable and sharp for more than 150 hours . 
As a result, we can make the molten zone smaller, which helps to keep the growth conditions stable. 
Such stability is required to grow large crystals with narrow mosaic spreads and small concentration gradients. 
The shapes of the resulting crystals are columnar, with typical dimensions of 7-8 mm in diameter and ~100 mm in length. 
Crystal rods near the final part of the growth were cut into $\sim$30 mm long pieces for neutron scattering measurements. 
All crystals were annealed under oxygen gas flow at 900 $^{\circ}$C for 50 hours, cooled to 500 $^{\circ}$C at a rate of 10 $^{\circ}$C/h, annealed at 500 $^{\circ}$C for 50 hours, and finally furnace-cooled to room temperature. 
The sample with {\it x} = 0.06 is the same crystal as that used for a previous neutron scattering study.~\cite{Waki_99_2}

The upper and lower parts of the crystals used for neutron scattering measurements were cut into $\sim$1 mm thick pieces in order to measure the superconducting shielding signal with a SQUID magnetometer. 
As shown in Fig. 1(a), superconducting transitions are observed in those samples with {\it x} = 0.056, 0.06, and 0.07, having onset temperatures {\it T$_{c}$} = 6.3 K, 11.6 K, and 17.0 K, respectively. 
These values are almost identical to those previously obtained on powder samples as shown in Fig. 1(b).~\cite{Takagi_89} 
The difference in {\it T$_{c}$} between the upper and lower parts of each crystal was found to be less than 0.1 K, which indicates the absence of any significant gradients in the Sr concentration {\it x}. 
On the other hand, no evidence for superconductivity is found for samples with {\it x} = 0.053 down to 2 K.  
Based on these results we estimate the lower critical concentration for superconductivity to lie between {\it x} = 0.053 and 0.056.

The orthorhombic distortion ({\it b}/{\it a}-1) measured at lowest temperature ($\sim$2 K) for all samples is shown in Fig. 2, along with data obtained on other crystals during our previous neutron diffraction measurements. 
We note that this distortion depends mainly on the Sr concentration, and is independent of the oxygen content. 
Therefore the linear fit to these data implies that we have achieved a systematic and well-defined Sr concentration in our samples. 
The slight scatter in the data is due primarily to error from experimental uncertainties caused by the use of different spectrometer configurations during different experiments. 
Similarly, both {\it T$_{c}$} and the lattice constants, which do depend on the oxygen content, can be used to gauge the oxygen stoichiometry. 
Fig. 1(b) shows a smooth and monotonic variation of {\it T$_{c}$} with {\it x}. 
Since the Sr content is known to vary systematically from our data on the 
\linebreak
\vspace{-0.5cm}
\begin{figure}[h]
\centerline{\epsfxsize=2.85in\epsfbox{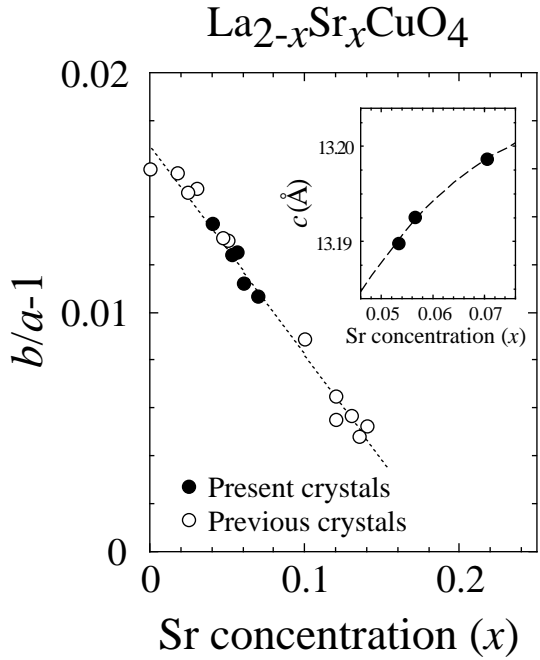}}
\vspace{0.5cm}
\caption
{Orthorhombic distortion ({\it b}/{\it a}-1) as a function of Sr concentration (solid circles) at lowest temperature together with previous results (open circles). {\it a} and {\it b} are the in-plane lattice constants for the orthorhombic lattice. Inset shows the {\it c}-axis lattice constant for {\it x} = 0.053, 0.056 and 0.07 at room temperature. Dashed lines are guides to the eye.}
\end{figure}
\noindent
orthorhombic distortion, the absence of any significant scatter in these data likewise suggests the absence of any scatter in the oxygen contents of these crystals. 
As an additional cross check, we measured the {\it c}-axis lattice constant for crystals with {\it x} = 0.053, 0.06, and 0.07 using an identical x-ray powder diffraction setup. 
Again, the {\it c}-axis lattice constant, shown in the inset to Fig. 2, also changed monotonically with Sr doping. 
Based on these results, and the agreement between current and prior results concerning the Sr-concentration dependence of {\it T$_{c}$}, we believe that the oxygen content in these crystals is stoichiometric. 
The maximum deviation of the Sr-concentration from the average value in each crystal is estimated to be $\sim$0.004, which is comparable to that previously evaluated for LSCO single crystals for 0.06 $\leq$ {\it x} $\leq$ 0.12, since the same growth techniques are utilized.~\cite{Yamada_98} 
The hole concentration is equal to the Sr concentration.

\subsection{Neutron Scattering Measurements}

The primary elastic neutron-scattering measurements reported here were performed on the cold neutron triple-axis spectrometers HER, located at the Japan Atomic Energy Research Institute (JAERI) JRR-3M reactor, and SPINS, located at the National Institute of Standards and Technology (NIST) Center for Neutron Research. 
Most experiments were carried out using incident and scattered neutron energies of 4.25 meV at HER, and 3.5 meV at SPINS, selected via Bragg diffraction from the (0,0,2) reflection from highly-oriented pyrolytic graphite crystals. 
Elastic scattering measurements were also performed on the thermal neutron triple-axis spectrometer TOPAN, located at JAERI. 
In this case the initial neutron energy was fixed at 14.7 meV.  Horizontal collimations used were 32$^{\prime}$-100$^{\prime}$-Be-S-80$^{\prime}$-80$^{\prime}$ at HER, 32$^{\prime}$-Be-80$^{\prime}$-S-BeO-80$^{\prime}$-150$^{\prime}$ at SPINS and 40$^{\prime}$-100$^{\prime}$-S-PG-60$^{\prime}$-80$^{\prime}$ at TOPAN. 
Here "S" denotes the sample position. 
Be, BeO and PG filters were used to eliminate contamination from higher-order wavelengths in the incident and scattered neutron beams. 
The resultant elastic energy resolution is about 0.2 meV and 2.0 meV FWHM (full-width at half-maximum) for cold and thermal neutron spectrometers, respectively. 
Crystals were mounted in the ({\it h},{\it k},0) zone, and sealed in an aluminum can with He gas for thermal exchange. 
The aluminum cans were then attached either to the cold plate of a $^{4}$He-closed cycle refrigerator, or to a top-loading liquid-He cryostat, which is able to control the temperature from 1.5 K to 300 K.

All crystals examined in this study have a twinned-domain orthorhombic structure and contain two types of domains. 
The volume ratio of the two domains is approximately 2:1. 
For samples with {\it x} = 0.053, 0.056, and 0.06, domain-selective scans were done as described in Ref. 11. 
It is convenient to use a polar coordinate {\it Mod}-{\it q}, measured in reciprocal lattice units (r.l.u.) of the high-temperature tetragonal structure (1 r.l.u. $\sim$ 1.65 ${\AA}^{-1}$), to describe the distance between incommensurate peaks, irrespective of the propagation vector. 
As illustrated in Fig. 3, the polar coordinate {\boldmath ${Mod}$}-{\boldmath ${q}$} = {\boldmath $Q$} - {\boldmath $Q$}$_{center}$, where {\boldmath $Q$}$_{center}$ is the vector from the 
\linebreak
\begin{figure}[hbt]
\centerline{\epsfxsize=2.6in\epsfbox{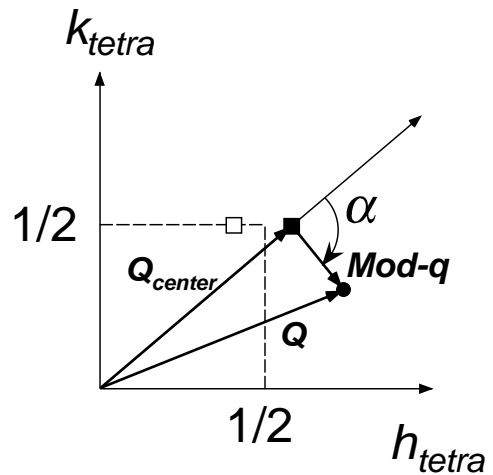}}
\vspace{0.5cm}
\caption
{Definition of the polar coordinates {\boldmath ${Mod}$}-{\boldmath ${q}$} and $\alpha$ centered at (1/2,1/2,0) for the two-dimensional tetragonal reciprocal lattice. Solid and open squares represent the orthorhombic (1,0,0) and (0,1,0) positions from the two domains.}
\end{figure}
\noindent
origin to the orthorhombic (1,0,0) position of the largest domain (represented by the solid square), and {\boldmath $Q$} is the momentum transfer. 
We define $\delta$, often called the incommensurability, as half the distance between the pair of IC peaks. 
The orientation of the peaks is described by a polar angle $\alpha$ that is measured with respect to {\boldmath $Q$}$_{center}$ and {\boldmath ${Mod}$}-{\boldmath ${q}$}. 
We note that this polar-geometry description of $\delta$ is also useful to describe the incommensurate 
peak positions observed in La$_{2}$CuO$_{4.12}$~\cite{YoungLee_99} and 
La$_{1.88}$Sr$_{0.12}$CuO$_{4}$.~\cite{Kimura_99} 
In this paper we present most of our {\it q}-scans using {\boldmath ${Mod}$}-{\boldmath ${q}$} as the horizontal axis.

\section{MAGNETIC SCATTERING CROSS SECTION}

In this section we present neutron elastic magnetic scattering data for samples with {\it x} = 0.04, 0.053 and 0.06. 
Fig. 4 shows the elastic scattering profiles of the magnetic peaks measured along the diagonal scan trajectory (see the top panel inset) for the insulating {\it x} = 0.04 and 0.053 samples at 1.5 K, along with prior results obtained for {\it x} = 0.05.~\cite{Waki_00_1} 
The elastic peaks observed in the insulating {\it x} = 0.04 and 0.053 samples are located at diagonal reciprocal lattice positions along the orthorhombic [0,1,0] direction, which is in complete agreement with the results obtained by Wakimoto {\it et al}. for {\it x} = 0.05.~\cite{Waki_00_1} 
By contrast, scans on the superconducting {\it x} = 0.06 sample (near {\it x$_{c}$}) indicate the presence of elastic peaks at parallel reciprocal lattice positions along the tetragonal [1,0,0] direction, which is consistent with previous results on superconducting samples~\cite{Waki_99_2} (Fig. 5(a)). 
To our surprise, however, we also observe {\it diagonal} elastic peaks at low temperatures for this same superconducting {\it x} = 0.06 sample as shown in Fig. 5(b). The data represented by the open circles in this panel were taken using the same scan at 40 K, and demonstrate that these diagonal peaks vanish at high temperatures, and are therefore genuine.

To clarify this observation we performed a circle scan denoted in the inset of Fig. 5(c), thereby obtaining a more detailed two-dimensional peak profile of these unexpected diagonal peaks. 
We note, as will be shown later, that the incommensurabilities of the peaks at the diagonal and parallel positions are nearly the same.  In other words, the diagonal and parallel peaks are nearly equidistant from the center of our polar coordinate system ({\boldmath ${Mod}$}-{\boldmath ${q}$} = 0), and thus all lie on a circle of radius $\delta$. 
Therefore, the circular scan is able to survey each of these peak profiles at once. 
As shown in Fig. 5(c), the circular scan reveals a single broad peak centered at $\alpha$ = 90$^{\circ}$ (see Fig. 3). 
From these scans, we can conclude that a pair of crescent-shaped peaks is present for {\it x} = 0.06, and is centered at the diagonal positions. A similar feature was observed in the {\it x} = 0.056 sample. We remark here that these measurements were performed by 
choosing magnetic peaks from the major domain, and are free from contamination by the corresponding peaks from the minor domain.
The 
\linebreak
\vspace{-0.8cm}
\begin{figure}[hbt]
\centerline{\epsfxsize=3in\epsfbox{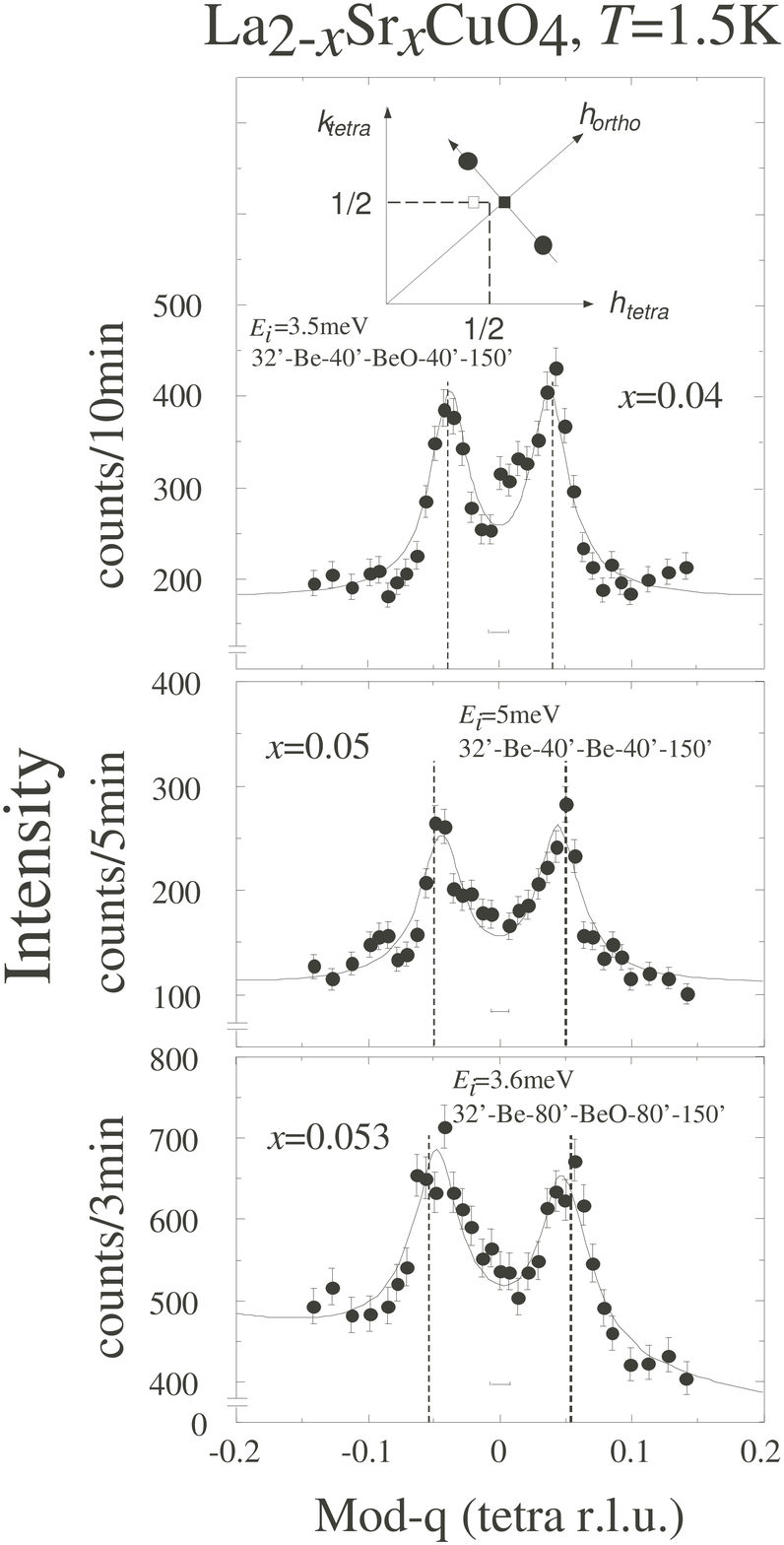}}
\vspace{0.5cm}
\caption
{Elastic peak profiles at 1.5 K along the diagonal scan direction for samples with {\it x} = 0.04, 0.05 and 0.053. The scattering profile for the {\it x} = 0.05 sample is taken from Ref. 10. The dashed lines represent the peak positions expected from the relation $\delta$ = {\it x}. The horizontal bars indicate the instrumental {\it q}-resolution FWHM.}
\end{figure}
\noindent
crescent-shaped peak revealed by three scans in Figs. 5(a)-(c) is quite different from the situation found for {\it x} = 0.05~\cite{Waki_99_1} or 0.12.~\cite{Kimura_99} 
In these two cases, a pair of nearly isotropic diagonal peaks was observed for the {\it x} = 0.05 sample, whereas four isotropic peaks were observed at parallel positions for the {\it x} = 0.12 sample. In order to analyze the data for {\it x} = 0.06, we used a model in which distinct, isotropic peaks were assumed to coexist at both diagonal and parallel positions, as shown in the inset of Fig. 5(c), hereafter referred to as the coexistence model. The fitting parameters were refined so as to reproduce three profiles shown in Figs. 5(a)-(c) simultaneously. 
\linebreak
\vspace{-0.8cm}
\begin{figure}[hbt]
\centerline{\epsfxsize=2.95in\epsfbox{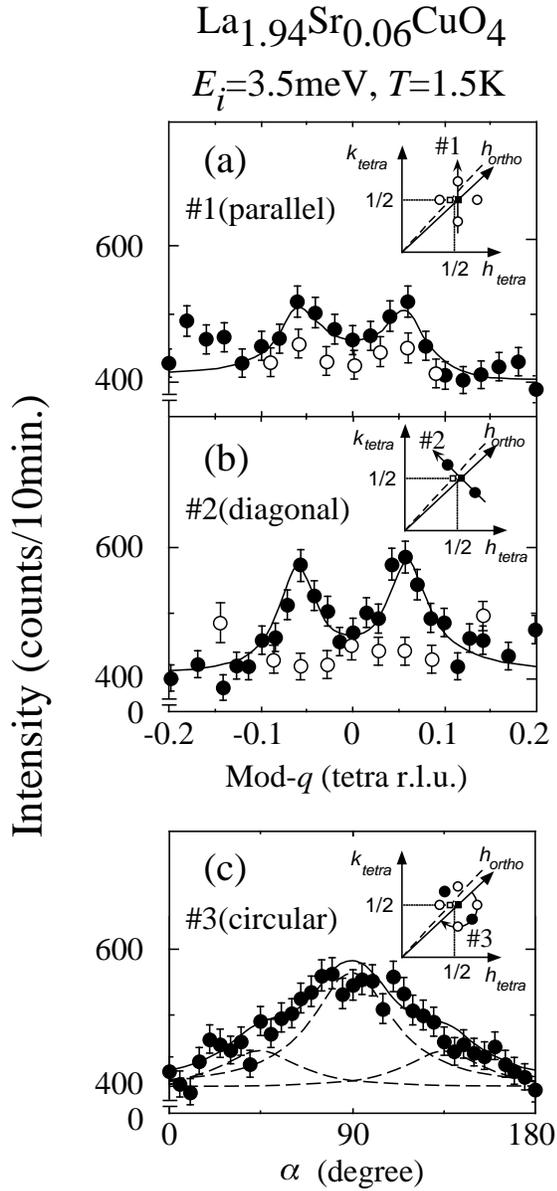}}
\vspace{0.62cm}
\caption
{Elastic magnetic peak profiles for {\it x} = 0.06 at 1.5 K (closed circles) and 40K (open circles) obtained by scan (a) ${\#}$1, (b) ${\#}$2, and (c) ${\#}$3, as shown in each inset. A solid (open) square denotes the orthorhombic (1,0,0) ((0,1,0)) position for the major (minor) domain. The solid lines in Figs. 5(a)-(c) are calculated assuming four peaks at parallel positions (open circles) and two peaks at diagonal positions (solid circles). Dashed lines in (c) represent the individual scattering contribution calculated for each peak.}
\end{figure}
\noindent
The obtained parameters for the diagonal peaks incommensurability $\delta$, and peak-width $\kappa$, are 0.053$\pm$0.002 (r.l.u.) and 0.039$\pm$0.004 ($\AA$$^{-1}$), respectively. 
The corresponding values for the parallel peaks are $\delta$ = 0.049$\pm$0.003 (r.l.u.) and $\kappa$ = 0.03$\pm$0.006 ($\AA$$^{-1}$). 
The resulting intensity ratio between the sum of the four parallel peaks and the two diagonal peaks is approximately 1:2. 
Note that in the {\it x} = 0.07 sample obvious IC peaks are observed at the parallel positions although the elastic magnetic signal is relatively weaker compared with the data for {\it x} = 0.06 samples and parallel component dominates the diagonal one.

Another new finding in this study is the observation of an anomalous dependence of the peak-width $\kappa$ on doping. 
In Fig. 6 we plot the doping dependence of $\kappa$, measured along the spin modulation vector, together with data from previous studies for both sides of the boundary.~\cite{Kimura_99}$^{,}$~\cite{Waki_00_1}$^{,}$~\cite{Matsuda_00}$^{,}$~\cite{Matsushita} 
A remarkable enhancement of $\kappa$ for both the diagonal and parallel elastic peaks are clearly observed in the superconducting phase near {\it x$_{c}$}. We remark that this enhancement is not simply caused by the overlap of two peaks because our analysis fits the width of each peak separately. 
We also note that a similar enhancement was already shown in Ref. 9 in which, however, the peak-width $\kappa$ was evaluated without domain-selective scans for {\it x} $\leq$0.05 and using low-energy inelastic signals for {\it x} $\geq$0.06. On the other hand, we show here the $\kappa$ using only elastic signals taken by domain-selective scans. 

Figs. 7(a) and (b) show the $\delta$ and peak-angle $\alpha$ (defined in Fig. 3) versus Sr concentration {\it x} for both the parallel (open circles) and diagonal (solid circles) elastic peaks. In Fig. 7(b), the size of the circles represents the relative intensities of the peaks, and the vertical bars correspond to the peak-width FWHM measured along the circle of 
\linebreak
\vspace{0cm}
\begin{figure}[hbt]
\centerline{\epsfxsize=2.75in\epsfbox{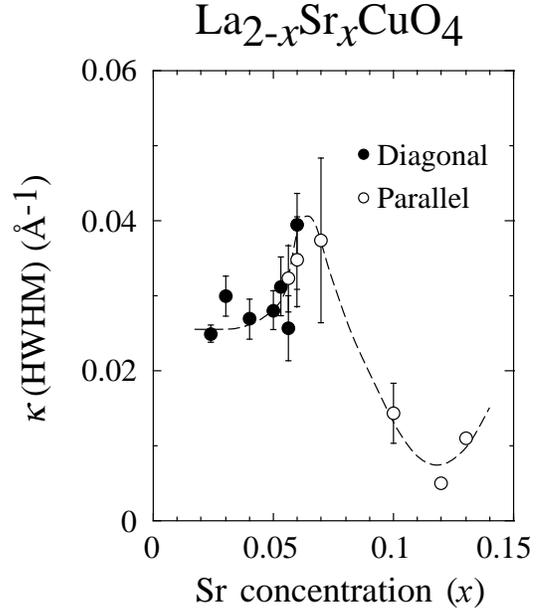}}
\vspace{0.5cm}
\caption
{Peak-width of the magnetic IC signal along the spin modulation vector as a function of Sr concentration. Open and solid circles represent the peak-width for the diagonal and parallel components, respectively. Data for {\it x} = 0.024,~\cite{Matsuda_00} 0.05,~\cite{Waki_00_1} 0.12,~\cite{Kimura_99} 0.1,~\cite{Matsushita} and 0.13~\cite{Matsushita} are also plotted in the figure.}
\end{figure}
\noindent
\linebreak
\vspace{-7mm}
\begin{figure}[hbt]
\centerline{\epsfxsize=2.9in\epsfbox{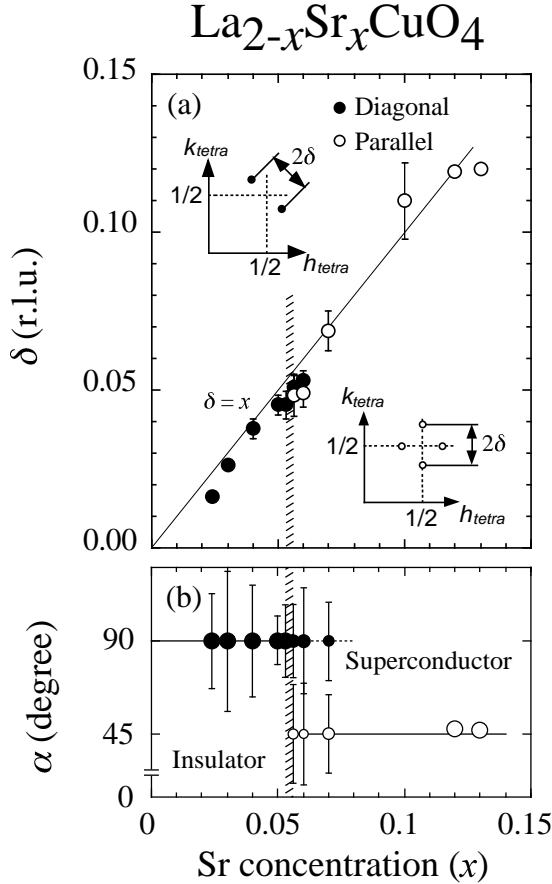}}
\vspace{0.44cm}
\caption
{Sr-concentration dependence of (a) the incommensurability $\delta$ and (b) the angle a defined in Fig. 3. Previous results for {\it x} = 0.024,~\cite{Matsuda_00} 0.04,~\cite{Waki_00_1} 0.05,~\cite{Waki_00_1} 0.12,~\cite{Kimura_99} 0.1,~\cite{Matsushita} and 0.13~\cite{Matsushita} are included. In both figures, the solid and open symbols represent the results for the diagonal and parallel components, respectively. }
\end{figure}
\noindent
radius $\delta$. For samples on which we did not perform any circular scans, peak-width in angle unit is calculated from that perpendicular to the spin modulation vector.  The value of $\delta$ for both peaks approximately follows the simple linear relation $\delta$ = {\it x}, except for {\it x} = 0.024 as is discussed in Ref. 11. 
However, as seen in Fig. 7(b), the intensity at the parallel positions appears beyond {\it x$_{c}$}, in accordance with the onset of the superconductivity. 
These results suggest that when the parallel component first appears at {\it x$_{c}$}, it does so with a finite incommensurability. 
Here we remark that the value of $\delta$ near the phase boundary exhibits no discontinuous change, but does show a slight downward deviation away from the relation $\delta$ = {\it x}.

\section{Discussion}

In this study we experimentally clarified how the change in the spin modulation vector takes place upon crossing the phase boundary at {\it x} = {\it x$_{c}$}. 
In the superconducting phase well-defined peaks appear at parallel positions, although the diagonal component seen in the spin-glass phase persists. 
The intensity of the parallel component becomes dominant with the development of superconductivity upon further doping. 
The coincidence of the parallel component and the superconductivity with Sr doping indicates an intimate connection between the parallel spin modulation and the superconductivity (or the itinerant holes on the antiferromagnetic Cu-O square lattices). 
Another important result is that the incommensurabilities for both diagonal and parallel peaks monotonously connect at {\it x} = {\it x$_{c}$} (see Fig. 7(a)). 
Hence it is revealed that over a wide range of Sr concentration {\it x}, the value of $\delta$ for both the diagonal and parallel components follows the linear relation $\delta$ = {\it x}, even though the spin modulation vectors for the two components are entirely different. 
Note that if two types of hole stripes coexist corresponding to the two types of spin modulations, the average hole density for each stripe phase is nearly the same at {\it x} = {\it x$_{c}$}. 

We believe that having different spin modulation vectors present on either side of the phase boundary is a key to understanding the nature of the doping-induced superconducting phase transition. 
As already mentioned in the previous section, the coexistence of isotropic diagonal and parallel peaks in the superconducting phase near {\it x$_{c}$} reproduces the observed spectra in Fig. 5 quite well. 
In the {\it x} = 0.06 sample, both insulating and superconducting phases may coexist or phase-separate either microscopically or mesoscopically, and our neutron scattering data strongly suggest the former (latter) phase is accompanied with the diagonal (parallel) spin modulation. 

We now briefly remark an alternative model to describe the drastic change in the spin modulation vector at {\it x} = {\it x$_{c}$}. In this model, it is assumed that a pair of diagonal peak splits continuously into four crescent-shaped peaks centered between the diagonal and parallel positions. 
Although this model also reproduces three profiles in Fig. 5, the fitted parameters for the incommensurability and the peak-width along the radial direction are approximately the same as those obtained by the coexistence model. 
Therefore, results for the incommensurability and the peak-width shown in Fig. 6 and Fig. 7(a) do not depend on the details of the change in the spin modulation vector near {\it x$_{c}$}. 
Moreover, if inhomogeneous hole-distribution exists in these samples by either extrinsic or intrinsic reason, it is very difficult to distinguish two models or two situations. 
The continuous change in the spin modulation vector near {\it x$_{c}$}, if it occurred, is difficult to detect experimentally. 
On the other hand, in general case with such inhomogeneous hole-distribution the continuous change turns into the two phase-mixing at the phase boundary. 
In the present system, we, therefore, expect two phase separation rather than continuous change of the direction of spin modulation vector. 
The anomalous broadening of peak-width is also consistent with the existence of two phases. 
If both phases microscopically phase-separate, then one phase will impede the other from expanding the size of its ordered regions upon cooling, and hence result in broadened peaks. 
Phase separation is also suggested by the incomplete Meissner effect observed near the phase boundary, and the subsequent increase of the superconducting volume fraction upon further doping.

Recall that the incommensurability $\delta$ of the parallel peak when it first appears is non-zero above {\it x$_{c}$}, and that the $\delta$ of the diagonal peak connects smoothly to that of the parallel one at {\it x$_{c}$}. 
Assuming that a proportional relationship exists between {\it T$_{c}$} and $\delta$ in the superconducting phase,~\cite{Yamada_98} a non-zero value for $\delta$ at {\it x$_{c}$} would imply a finite value for {\it T$_{c}$} at {\it x$_{c}$}. 
Within our experimental uncertainties for {\it T$_{c}$}, this conclusion is not inconsistent with the experimental results presented in Fig. 1(b). Previously, in the hole-doped high-{\it T$_{c}$} cuprates, the transition temperature {\it T$_{c}$} has been considered to decrease continuously down to zero with decreasing hole content. 
However, the present neutron scattering results strongly suggest the discontinuous change in {\it T$_{c}$} at the phase boundary. Such a first-order transition of superconductivity at the boundary is supported by the result of recent resistivity measurement.~\cite{Fujita_00}

In conclusion, we confirmed the coincident appearance of the parallel spin modulation and the superconductivity upon hole-doping. The diagonal spin modulation persists into the superconducting state near the phase boundary. The incommensurabilities for the diagonal and parallel spin modulations monotonously connect at the boundary.

\section*{Acknowledgments}
We would like to thank R. J. Birgeneau, V. J. Emery, Y. Endoh, K. Hirota, K. Kimura, K. Machida, M. Matsuda, J. Tranquada, and Y. S. Lee for valuable discussions. We also would like to acknowledge Y. Ikeda for his assistance with the sample characterization. This work was supported in part by the Japanese Ministry of Education, Culture, Sports, Science and Technology, Grant-in-Aid for Scientific Research on Priority Areas (Novel Quantum Phenomena in Transition Metal Oxides), 12046239, 2000, for Scientific Research (A), 10304026, 2000, for Encouragement of Young Scientists, 13740216, 2001 and for Creative Scientific Research (13NP0201) "Collaboratory on Electron Correlations - Toward a New Research Network between Physics and Chemistry -", by the Japan Science and Technology Corporation, the Core Research for Evolutional Science and Technology Project (CREST). Work at Brookhaven National Laboratory was carried out under Contract No. DE-AC02-98-CH10886, Division of Material Science, U. S. Department of Energy. We also acknowledge the support of the National Institute of Standards and Technology (NIST), U.S. Department of Commerce, in providing the neutron facilities used in this work.  Work performed on SPINS spectrometer in NIST was also based upon activities supported by the National Science Foundation under Agreement No. DMR-9423101. The work at MIT was supported by the NSF under Grant No. DMR0071256 and by the MRSEC Program of the National Science Foundation under Grant No. DMR98-08941.




\begin{references} 
\vspace{-1.0cm}
%
\bibitem[\ast] {} Now at the Institute for Material Research, Tohoku University, Katahira, Sendai 980-0821.
%
\bibitem[\dagger] {} Also at the University of Maryland, College Park, MD 20742
%
\bibitem[\ddagger] {} Massachusetts Institute of Technology, Cambridge, MA 02139.  Present address: Department of Physics, University of Toronto, Toronto, Ontario, Canada M5A 1A7. 
%
\bibitem[1] {Kastner_98} M. A. Kastner, R. J. Birgeneau, G. Shirane, and Y. Endoh, 
Rev.\ Mod.\ Phys. {\bf 70}, 897 (1998).
\bibitem[2] {Yamada_98} K. Yamada, C. H. Lee, K. Kurahashi, J. Wada, S. Wakimoto, 
S. Ueki, H. Kimura, Y. Endoh, S. Hosoya, G. Shirane, R.J. Birgeneau, M. Greven, 
M.A. Kastner, and Y.J. Kim, 
Phys.\ Rev.\ B {\bf 57}, 6165 (1998). 
\bibitem[3] {Tranquada_95} J. M. Tranquada, B. J. Sternlieb, J. D. Axe, Y. Nakamura, and S. Uchida, Nature (London) {\bf 375}, 561 (1995).
\bibitem[4] {Suzuki_98} T. Suzuki, T. Goto, K. Chiba, T. Shinoda, T. Fukase, H. Kimura, K. Yamada, M. Ohashi, and Y. Yamaguchi, Phys. Rev. B {\bf 57}, 3229 (1998).
\bibitem[5] {Kimura_99} H. Kimura, K. Hirota, H. Matsushita, K. Yamada, Y. Endoh, S. H. Lee, C. F. Majkrzak, R. Erwin, G. Shirane, M. Greve, Y. S. Lee, M. A. Kastner, and R. J. Birgeneau, Phys. Rev. B {\bf 59}, 6517 (1999).
\bibitem[6] {YoungLee_99} Y. S. Lee, R. J. Birgeneau, M. A. Kastner, Y. Endoh, S. Wakimoto, K. Yamada, R. W. Erwin, S. H. Lee, and G. Shirane, Phys. Rev. B {\bf 60}, 3643 (1999).
\bibitem[7] {Tranquada_96} J. M. Tranquada, J. D. Axe, N. Ichikawa, Y. Nakamura, S. Uchida, and B. Nachumi, Phys, Rev. B {\bf 54}, 7489 (1996).
\bibitem[8] {Kimura_00} H. Kimura, H. Matsushita, K. Hirota, Y. Endoh, K. Yamada, G. Shirane, Y. S. Lee, M. A. Kastner, and R. J. Birgeneau, Phys. Rev. B {\bf 61}, 14366 (2000).
\bibitem[9] {Waki_99_1} S. Wakimoto, G. Shirane, Y. Endoh, K. Hirota, S. Ueki, K. Yamada, R. J. Birgeneau, M. A. Kastner, Y. S. Lee, P. M. Gehring, and S. H. Lee, Phys. Rev. B {\bf 60}, R769 (1999).
\bibitem[10] {Waki_00_1} S. Wakimoto, R. J. Birgeneau, M. A. Kastner, Y. S. Lee, R. Erwin, P. M. Gehring, S. H. Lee, M. Fujita, K. Yamada, Y. Endoh, K. Hirota, and G. Shirane, Phys. Rev. B {\bf 61}, 3699 (2000).
\bibitem[11] {Matsuda_00} M. Matsuda, M. Fujita, K. Yamada, R. J. Birgeneau, M. A. Kastner, Y. Endoh, S. Wakimoto, and G. Shirane, Phys. Rev. B {\bf 62}, 9148 (2000).
\bibitem[12] {Tanaka_89} I. Tanaka, K. Yamane, and H. Kojima, J. Crystal Growth {\bf 96}, 711 (1989).; S. Hosoya, C. H. Lee, S. Wakimoto, K. Yamada, and Y. Endoh, Physica C 235-240, {\bf 547} (1994).;
C. H. Lee, N. Kaneko, S. Hosoya, K. Kurahashi, S. Wakimoto, K. Yamada, and Y. Endoh, Supercond. Sci. Technol. {\bf 11}, 891 (1998).
\bibitem[13] {Waki_99_2} S. Wakimoto, K. Yamada, S. Ueki, G. Shirane, Y. S. Lee, 
S. H. Lee, M. A. Kaster, K. Hirota, P. M. Gehring, Y. Endoh, and R. J. Birgeneau, J.Phys. Chem. Solid, {\bf 60}, 1079 (1999).
\bibitem[14] {Takagi_89} H. Takagi, T. Ido, S. Ishibashi, M. Uota, and S. Uchida, Phys. Rev. B {\bf 40}, 2254 (1989).; H. Takagi, B. Batlogg. H. L. Kao, J. Kwo, R. J. Cava, J. J. Krajewski, and W. F. Peck, Jr, Phys. Rev. Lett. {\bf 69}, 2975 (1992).
%
\vspace{30cm}
\bibitem[15] {Matsushita} H. Matsushita, H. Kimura, M. Fujita, K. Yamada, K. Hirota, and Y. Endoh, J. Phys. Chem. Solid {\bf 60}, 1071 (1999).
\bibitem[16] {Waki_00_2} S.Wakimoto, S.Ueki, Y.Endoh, and K.Yamada, Phys. Rev. B {\bf 62}, 3547 (2000).
\bibitem[17] {Fujita_00} T. Fujita, J. Hori, T. Goko, N. Kikugawa, and S. Iwata, Inst. Phys. Chem. Res. {\bf 27}, 75 (2000).

\end{references}
\end{document}